\documentclass[11pt]{article}

\usepackage[final]{acl}
\usepackage{times}
\usepackage{latexsym}

\usepackage[T1]{fontenc}

\usepackage[utf8]{inputenc}

\usepackage{microtype}

\usepackage{inconsolata}

\usepackage{graphicx}
\usepackage{amsmath}   %
\usepackage{amssymb}   %
\usepackage{amsfonts}  %
\usepackage{amsthm}
\newtheorem{definition}{Definition}
\usepackage{amsbsy}    %
\usepackage{todonotes}

\usepackage[ruled,linesnumbered]{algorithm2e}
\usepackage{subcaption}
\usepackage{array,booktabs,multirow}
\usepackage{colortbl}
\usepackage{threeparttable}
\usepackage{float}
\newcommand{\se}[1]{\textcolor{gray}{\scriptsize #1}}

\usepackage{pifont}
\newcommand{\cmark}{\ding{51}}
\newcommand{\xmark}{\ding{55}}

\usepackage[most]{tcolorbox} %
\tcbuselibrary{listings, breakable}
\usepackage{cleveref}

\newcounter{querybox}
\crefname{querybox}{Box}{Boxes}
\Crefname{querybox}{Box}{Boxes}

\newtcolorbox{queryboxenv}[1][]{%
  enhanced, breakable,
  colback=white, colframe=black,
  colbacktitle=black, coltitle=white, fonttitle=\bfseries,
  sharp corners, boxrule=0.6pt,
  left=1em, right=1em, top=0.75em, bottom=0.75em,
  before upper=\refstepcounter{querybox}, %
  #1 %
}

\title{An Empirical Study on Zero-Data Bootstrapping for \\
Conversational Recommender Systems}

\author{
\textbf{Rohan Surana$^{1,*}$, Junda Wu$^{1,*}$, Zhouhang Xie$^{1,*}$, Yu Xia$^{1}$, Nathan Kallus$^{2,3}$, Julian McAuley$^{1}$}\\
$^{1}$University of California, San Diego \\
$^{2}$Netflix Inc. $^{3}$Cornell University \\
\texttt{\{rsurana, juw069, zhx022, yux078, jmcauley\}@ucsd.edu} \\
\texttt{\{nkallus\}@netflix.com} \\
}
\begin{document}
\maketitle

\begin{abstract}
Conversational Recommender Systems (CRS) typically require domain-specific dialogue data, which is costly, scarce, and often unavailable in new domains.
We conduct a systematic empirical study of \emph{zero-data CRS bootstrapping}: generating synthetic conversational supervision from non-conversational signals---item reviews, metadata, and user-item interactions---without any in-domain dialogue corpus.
We compare two information-theoretic selection strategies, Jensen-Shannon diversity and Fisher information, across domain signals, model architectures, datasets, and fine-tuning paradigms.
Our results show that domain-grounded synthetic data consistently outperforms zero-shot prompting and na\"ive synthetic baselines; active selection improves data efficiency over random sampling; metadata and collaborative filtering signals each improve selection quality; and, in low-resource settings, synthetic data can outperform scarce real dialogues while further complementing them.
These findings establish non-conversational domain signals as a viable path toward building CRS without conversational training data.
The code is available at \url{https://anonymous.4open.science/r/zero_data_crs/}.
\end{abstract}

\section{Introduction}

\begin{figure}[htb]
    \centering
    \includegraphics[width=\linewidth]{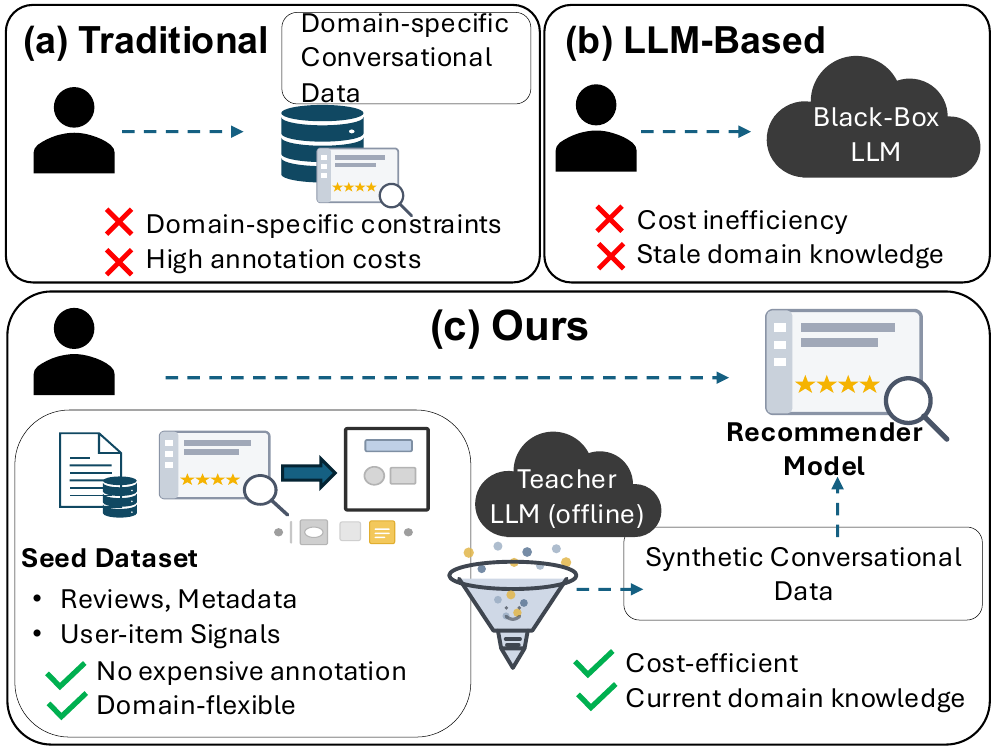}
    \caption{Three CRS paradigms: (a) traditional CRS requiring domain-specific conversational data, (b) direct LLM deployment with inherent limitations, and (c) the zero-data bootstrapping pipeline studied in this work, which synthesizes conversational data from non-conversational seed signals.}
    \label{fig:fig1}
\end{figure}

Conversational Recommender Systems (CRS) aim to deliver personalized recommendations through dialogue~\citep{sun2018conversational,lin2023enhancing,wang2023target,xi2024memocrs,zhang2018towards}.
Building effective CRS has traditionally required large, domain-specific conversational datasets~\citep{wang22unicrs,chen-etal-2019-towards,zhou2020improving,li2018towards} (\Cref{fig:fig1}(a)), which are scarce due to \emph{high annotation costs}~\citep{soudani2024survey,chen2023places,aroyo2023dices}, \emph{privacy concerns}~\citep{reddit_api_update,reddit_pushshift_changes,reddit_public_content_policy,chen2023places}, and \emph{domain-specific constraints}~\citep{chen-etal-2019-towards,zhou2020improving,wang22unicrs}.
While Large Language Models (LLMs) demonstrate promising zero-shot capabilities~\citep{he2023large,feng2023large,friedman2023leveraging,wu2024survey}, direct deployment (\Cref{fig:fig1}(b)) faces challenges in \emph{scalability}~\citep{menshawy2024navigating,chen2025empirical}, \emph{cost}~\citep{samsi2023words,chen2023frugalgpt}, \emph{interpretability}~\citep{singh2024rethinking,kasneci2023chatgpt,yeo2024interpretable}, and \emph{privacy}~\citep{yao2024survey,neel2023privacy,yan2024protecting,li2024llm}, leading many practical deployments to prefer smaller, internally managed models~\citep{jeong2024fine}. Yet fine-tuning these models still requires substantial domain-specific conversational data~\citep{jeong2024fine,sun2024dial,nguyen2024enhancing}, which remains limited~\citep{soudani2024survey,wang2023improving1,ruiz2024fine}.

In parallel, LLMs have been increasingly used as data generators, synthesizing training data for downstream tasks~\citep{zhao2023alleviating,mysore2023large,leszczynski2023talk}. However, existing augmentation approaches for CRS operate \emph{on top of} an existing conversational corpus~\citep{xu-etal-2025-beyond,10.1145/3712588,zhao2023alleviating} or require external knowledge graphs~\citep{wang22unicrs,chen-etal-2019-towards,zhou2020improving}. Active learning offers principled sample selection to maximize the value of costly LLM queries~\citep{du2024active,liu2024evolving,zhang2024elad}, but its application to CRS data generation from non-conversational signals has not been studied.

This raises a fundamental question: \emph{can LLMs bootstrap effective CRS from non-conversational domain signals alone?} Non-conversational data---item reviews, metadata, and user-item collaborative interactions---is readily available in most domains, yet converting such signals into effective conversational training data involves non-trivial challenges in sample selection, domain grounding, and data efficiency. No systematic investigation exists of whether and how this bootstrapping can work.

In this work, we conduct a comprehensive empirical study addressing this question. We formalize the \emph{zero-data CRS} setting (\Cref{fig:fig1}(c)), where no in-domain conversational data is available, and study a three-stage bootstrapping pipeline: (i) active sample selection from non-conversational seed data using information-theoretic criteria, (ii) synthetic conversational data generation via a teacher LLM, and (iii) fine-tuning a target model on the synthesized corpus. We systematically vary domain signals, selection strategies, budgets, model architectures, and fine-tuning paradigms to address four research questions.

\paragraph{Contributions.}
We summarize our contributions as follows:
\begin{itemize}
    \item We formalize the \emph{zero-data CRS} setting and establish a controlled experimental framework for studying synthetic data bootstrapping from non-conversational domain signals.

    \item We provide a systematic empirical study of sample selection strategies for augmenting CRS data generation, characterizing how different selection criteria affect synthetic-data quality and downstream recommendation performance.

    \item We conduct controlled ablations isolating the effects of domain signals (item metadata, collaborative filtering) on both selection quality and downstream CRS performance across multiple models, datasets, and fine-tuning paradigms.
\end{itemize}

\paragraph{Scope.}
We focus on the strict setting where neither an in-domain conversational corpus nor an external knowledge graph is available. The only assumed resources are non-conversational domain signals such as reviews, metadata, and user-item interactions. \Cref{tab:setting_comparison} contrasts this scope with representative CRS and augmentation methods.

\begin{table}[t]
\scriptsize
\centering
\begin{tabular}{lccc}
\toprule
\textbf{Method} & \textbf{CRS Data} & \textbf{KG} & \textbf{Zero-Data} \\
\midrule
UniCRS~\citep{wang22unicrs} & \cmark & \cmark & \xmark \\
KGSF~\citep{zhou2020improving} & \cmark & \cmark & \xmark \\
KBRD~\citep{chen-etal-2019-towards} & \cmark & \cmark & \xmark \\
\citet{xu-etal-2025-beyond} & \cmark & \xmark & \xmark \\
NR-CRS~\citep{10.1145/3712588} & \cmark & \xmark & \xmark \\
LOT-CRS~\citep{zhao2023alleviating} & \cmark & \xmark & \xmark \\
\textbf{This Study} & \xmark & \xmark & \cmark \\
\bottomrule
\end{tabular}
\caption{Comparison of CRS methods by data requirements. CRS Data: requires in-domain conversational corpus. KG: requires external knowledge graph. Zero-Data: operates without conversational training data.}
\label{tab:setting_comparison}
\end{table}

\section{Zero-Data CRS Bootstrapping}
\label{sec:formulation}

\begin{figure*}[htp]
    \centering
    \includegraphics[width=\textwidth]{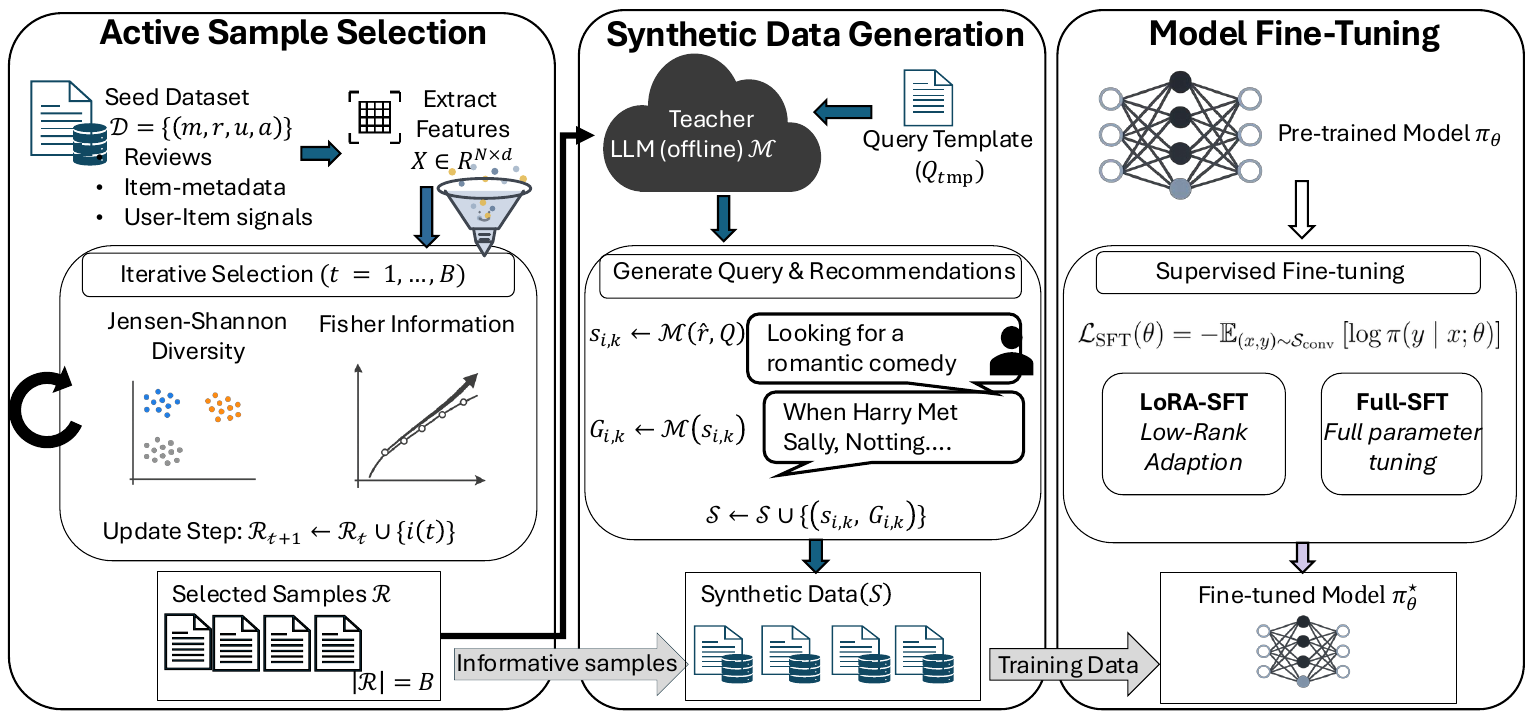}
    \caption{Overview of zero-data CRS bootstrapping. Non-conversational seed signals are encoded, actively selected, converted into synthetic conversations by a teacher LLM, and used to fine-tune the target CRS.}
    \label{fig:fig2}
\end{figure*}

We define the zero-data CRS setting and the bootstrapping pipeline evaluated in this study. \Cref{fig:fig2} provides an overview.

\subsection{Setting: Zero-Data CRS}

\begin{definition}[Zero-Data CRS]
\label{def:zero-data}
A CRS operates in the \emph{zero-data} setting when no in-domain conversational dataset $\mathcal{S}_{\mathrm{conv}}$ is available. The only available resources are non-conversational domain signals: item reviews $\{r_i\}$, item metadata $\{m_i\}$, and user-item collaborative interactions $\{u_i\}$.
\end{definition}

This setting captures the practical reality of deploying CRS in new or specialized domains where conversational data has never been collected, requiring conversational supervision to be derived from non-conversational signals alone.

\subsection{Pipeline: Select, Generate, Fine-Tune}
\label{sec:pipeline}

To study how non-conversational signals can be converted into effective CRS training data, we instantiate a three-stage bootstrapping pipeline (\Cref{alg:active_interface}). Given seed data $\mathcal{D}$, target model $\pi_\theta$, teacher LLM $\mathcal{M}$, and budget $B$, the pipeline selects informative samples $\mathcal{R}\subset\mathcal{D}$, generates synthetic conversations $\mathcal{S}$ from $\mathcal{R}$, and fine-tunes $\pi_\theta$ on $\mathcal{S}$ to obtain $\pi_\theta^*$.

\begin{algorithm}[t]
\caption{Zero-Data CRS Bootstrapping Pipeline}
\label{alg:active_interface}
\SetKwInOut{Input}{Input}\SetKwInOut{Output}{Output}
\Input{Seed dataset $\mathcal{D} = \{(m_i, r_i, u_i, a_i)\}_{i=1}^{N}$;
  Language model $\pi_\theta$; Black-box LLM $\mathcal{M}$;
  Query templates $\mathcal{Q}_{\mathrm{tmp}}$; Budget $B$}
\Output{Fine-tuned model $\pi_\theta^*$}
\BlankLine
\textbf{Initialize:} $\mathcal{S} \gets \emptyset$\;
\BlankLine
\ForEach{$i \in \mathcal{D}$}{
    $\mathbf{x}_i \gets \texttt{embed}(\pi_\theta, i)$
}
\BlankLine
$\mathcal{R} \gets \texttt{select}(\mathcal{D}, \{\mathbf{x}_i\}, B)$ \hfill $\triangleright$ \S\ref{sec:selection_strategies}
\BlankLine
$\mathcal{S} \gets \texttt{generate}(\mathcal{R}, \mathcal{Q}_{\mathrm{tmp}}, \mathcal{M})$ \hfill $\triangleright$ Alg.~\ref{alg:synthetic_data_creation}
\BlankLine
$\pi_\theta^* \gets \texttt{fine\_tune}(\pi_\theta, \mathcal{S})$ \hfill $\triangleright$ \S\ref{sec:finetuning}
\BlankLine
\Return{$\pi_\theta^*$}
\end{algorithm}

\subsection{Selection: Informative Seed Samples}
\label{sec:selection_strategies}

Given the seed dataset $\mathcal{D}=\{(m_i,\,r_i,\,u_i,\,a_i)\}_{i=1}^{N}$ with $N$ samples, where $m_i$ denotes item metadata, $r_i$ item reviews, $u_i$ user features derived from collaborative signals, and $a_i$ the item identifier, the goal is to select a subset of size $B$ that maximizes the quality of downstream synthetic data.

We extract dense representations by leveraging $\pi_\theta$'s last transformer layer hidden states $X \in \mathbb{R}^{N \times d}$, following prior work on LLM-based embeddings~\citep{tang2024pooling,wang2023improving2,li2024your,jiang2023scaling} to prevent sample misspecification~\citep{sugiyama2005active,fudenberg2017active,lin2023optimal}.

Selection precedes generation, so no training targets exist yet and label-dependent criteria such as uncertainty sampling do not apply. We therefore build on two label-free principles from active learning, diversity/coverage and Fisher-information-based optimal design~\citep{sener2018active,ash2021gone,mukherjee2024optimal,kveton2025active}, instantiated over $X$. The two criteria encode competing hypotheses about what makes a seed sample useful for CRS data generation: JS diversity prioritizes distributional coverage, while Fisher information prioritizes parameter-level informativeness under a last-layer approximation.

\begin{definition}[JS Diversity Selection]
\label{def:js_selection}
Let $p_i$ denote the soft cluster-membership distribution for sample $i$, obtained by clustering the embedding space and applying a softmax over distances to cluster centroids. Let $H(p_i)$ be its Shannon entropy. The first sample is initialized as the maximum-entropy item, after which $\bar{p}_t=\frac{1}{|\mathcal{R}_t|}\sum_{j\in\mathcal{R}_t}p_j$ is computed over the non-empty selected set. JS diversity then selects the next sample by
\begin{equation}
\label{eq:js_selection}
i^{(t)} =\arg\max_{i\in\mathcal{D}\setminus\mathcal{R}_t}
\Bigl[ H(p_i) + \lambda\,\operatorname{JS}\!\bigl(p_i \,\|\, \bar{p}_t\bigr) \Bigr].
\end{equation}
The entropy term favors boundary samples, while the JS term promotes novelty relative to the selected set.
\end{definition}

\begin{definition}[Fisher Information Selection]
\label{def:fisher_selection}
Let $\Lambda_t=\bigl(I_d+\sum_{j\in\mathcal{R}_t}\mathbf{x}_j\mathbf{x}_j^\top\bigr)^{-1}$ denote the inverse regularized information matrix for the current selected set. Under a last-layer linearization of the fine-tuning objective~\citep{mukherjee2024optimal,kveton2025active}, Fisher selection chooses the sample with the largest incremental information gain:
\begin{equation}
\label{eq:fisher_selection}
    i^{(t)} = \arg\max_{i\in \mathcal{D} \setminus \mathcal{R}_t}
    \log\!\Bigl(1 + \mathbf{x}_i^\top \Lambda_t \mathbf{x}_i\Bigr).
\end{equation}
This greedy log-determinant objective favors samples that add new parameter-relevant directions beyond those already selected.
\end{definition}

Together, the two criteria operationalize complementary notions of sample utility: distributional representativeness and parameter-aligned informativeness~\citep{kirsch2022unifyingapproachesactivelearning}. Fisher selection corresponds to greedy D-optimal design~\citep{doi:10.1137/1.9780898719109}, while JS selection is motivated by submodular coverage objectives~\citep{10.1007/BF01588971}. Both criteria are applied iteratively until $|\mathcal{R}|=B$; construction and update details are provided in \Cref{app:selection_criteria}.

\subsection{Pipeline: From Domain Signals to CRS Training Data}
\label{sec:transformation}

\paragraph{Domain Signal Encoding.}
\label{sec:domain_signals}

\begin{definition}[Domain Signals]
\label{def:domain_signals}
We distinguish three types of non-conversational domain signals available for seed sample representation:
\begin{itemize}
    \item \textbf{Semantic signals} ($r_i$): item review text, capturing fine-grained user opinions and item characteristics;
    \item \textbf{Structural signals} ($m_i$): item metadata including descriptions, genres, and categories, providing categorical and descriptive context;
    \item \textbf{Relational signals} ($u_i$): user-item collaborative filtering features derived from interaction histories, encoding user preference patterns.
\end{itemize}
\end{definition}

Each signal type is encoded into the embedding space by concatenating it with the seed input before extracting representations via $\pi_\theta$. This study systematically investigates how each signal type, and their combinations, affects both selection quality and downstream CRS performance.

\paragraph{Synthetic Conversation Generation.}
\label{sec:synthetic_data}
After active sample selection, the selected samples $\mathcal{R}$ are converted into synthetic conversational data using a teacher LLM. Let $\mathcal{Q}_{\mathrm{tmp}}$ be a collection of style templates sourced from Reddit movie recommendation discussions, used only to control the style and tone of generated queries (see \Cref{box:query_templates}).

For each selected item $i \in \mathcal{R}$, we sample five style templates $Q \subset \mathcal{Q}_{\mathrm{tmp}}$. For each of $K$ synthetic queries, we sample three reviews $\hat{r} \subset r_i$ and prompt a teacher LLM $\mathcal{M}$ to generate a synthetic query $s_{i,k}$ (\Cref{box:query}). Following~\citet{he2023large}, we then prompt the LLM with each query to produce 20 pseudo-target recommendations, $G_{i,k} = \{g_1, \dots, g_{20}\}$ (\Cref{box:resp}). Each pair $(s_{i,k}, G_{i,k})$ is added to the synthetic dataset $\mathcal{S}$ for fine-tuning the target language model $\pi_\theta$. For $B$ selected items and $K$ queries per item, this yields $BK$ training pairs at a cost of $2BK$ offline teacher calls; selection itself requires one embedding pass and no additional teacher calls. The full procedure is detailed in \Cref{alg:synthetic_data_creation}, with representative examples in \Cref{box:sample}.

\begin{algorithm}[t]
\caption{Synthetic conversational data generation from actively selected non-conversational seed items.}
\label{alg:synthetic_data_creation}
\SetKwInOut{Input}{Input}\SetKwInOut{Output}{Output}
\Input{Selected samples $\mathcal{R} \subset \mathcal{D}$; Query templates $\mathcal{Q}_{\mathrm{tmp}}$; Teacher LLM $\mathcal{M}$; Queries per item $K$}
\Output{Synthetic dataset $\mathcal{S} = \{(s, G)\}$, where $G$ contains pseudo-target recommendations}
\BlankLine
\textbf{Initialize:} $\mathcal{S} \gets \emptyset$\;
\BlankLine
\ForEach{$i \in \mathcal{R}$}{
    Sample 5 style templates $Q \subset \mathcal{Q}_{\mathrm{tmp}}$\;
    \For{$k = 1$ \KwTo $K$}{
        Sample 3 reviews $\hat{r} \subset r_i$\;
        $s_{i,k} \gets \mathcal{M}(\hat{r}, Q)$ \hfill $\triangleright$ Generate query\;
        $G_{i,k} \gets \mathcal{M}(s_{i,k})$ \hfill $\triangleright$ Generate 20 recommendation targets\;
        $\mathcal{S} \gets \mathcal{S} \cup \{(s_{i,k}, G_{i,k})\}$\;
    }
}
\BlankLine
\Return{$\mathcal{S}$}
\end{algorithm}

\paragraph{Target Model Fine-Tuning.}
\label{sec:finetuning}
We adapt $\pi_\theta$ to the conversational recommendation domain using supervised fine-tuning on $\mathcal{S}$, evaluating both LoRA and Full-SFT following standard practice~\citep{chen2024self, li2024getting, dong2023abilities,zhang2023instruction}. Training details and hyperparameters are provided in \Cref{impl_det}.

\section{Experimental Setup}
\label{sec:study_design}

\subsection{Benchmarks and Seed Data}
\label{sec:datasets}

We evaluate on two standard CRS benchmarks of different scales: ReDial~\citep{li2018towards} and INSPIRED~\citep{hayati2020inspired}. This pairing enables studying bootstrapping effectiveness across both moderate- and low-resource evaluation settings. Our non-conversational seed data is drawn from Amazon Reviews '23~\citep{hou2026bridginglanguageitemsretrieval} (Movies \& TV category). Dataset statistics are reported in \Cref{tab:dataset_stats}.

\begin{table}[ht]
\small
\centering
\begin{tabular}{lcc}
\toprule
 & ReDial & INSPIRED \\
\midrule
Total dialogues & 10,006 & 1,001 \\
Test dialogues & 2,001 & 200 \\
\bottomrule
\end{tabular}
\caption{Dataset statistics for the ReDial and INSPIRED evaluation benchmarks.}
\label{tab:dataset_stats}
\end{table}

\subsection{Backbone and Teacher Models}
\label{sec:backbone_models}

We study instruction-tuned LLMs as backbone models: Llama3.2-3B-Instruct~\citep{grattafiori2024llama}, Qwen2.5-1.5B-Instruct~\citep{qwen2.5}, and Qwen3-4B~\citep{qwen3technicalreport} (RQ1 only). This range of model sizes enables studying how model capacity interacts with synthetic data quality.

For fine-tuning, we consider two paradigms: (1) LoRA~\citep{hu2022lora}, parameter-efficient fine-tuning using Low-Rank Adaptation, and (2) Full-SFT, where all model parameters are optimized. The teacher LLM for synthetic data generation is GPT-4o~\citep{hurst2024gpt}. Implementation details are provided in \Cref{impl_det}.

\subsection{Baselines and Selection Conditions}
\label{sec:baselines}

To isolate the effect of domain-grounded synthetic data, we compare against baselines operating under the same zero-data constraint:

\noindent\textbf{Zero-Shot}: Models are directly prompted for conversational recommendation without fine-tuning.
\textbf{GPT-Generated}: Models fine-tuned on synthetic data generated by na\"ively prompting GPT-4o~\citep{hurst2024gpt} without domain-specific seed data or active selection.
\textbf{Popularity}: Recommends items based on frequency in the dataset.
\textbf{NBCRS}~\citep{xie2024neighborhood}: A neighborhood-based CRS evaluated both on raw seed data and on synthesized conversational data.

Existing CRS augmentation methods~\citep{xu-etal-2025-beyond,10.1145/3712588} require an in-domain conversational corpus and are therefore not applicable in the zero-data setting.

For active selection, we compare \textbf{Random} uniform sampling; semantic-only \textbf{JS}/\textbf{Fisher} selection using review embeddings; \textbf{JS+Meta}/\textbf{Fisher+Meta}, which add item metadata; and \textbf{JS+CF}/\textbf{Fisher+CF}, which add collaborative filtering (CF) signals from user interaction histories. We evaluate with Recall@$k$ ($k \in \{1, 5, 10, 20\}$) and NDCG@$k$ (\Cref{app:ndcg}); unless otherwise stated, RQ1--RQ3 train exclusively on synthetic data, while RQ4 studies combinations of synthetic and in-domain data.

\section{Main Results}

We organize the study around four questions: whether domain-grounded synthetic data improves zero-data CRS (RQ1), whether active selection improves data efficiency (RQ2), how metadata and collaborative signals affect selection (RQ3), and how synthetic data compares with scarce real dialogues (RQ4).

\subsection{RQ1: Domain-Grounded Synthetic Data Enables Zero-Data CRS}
\label{sec:rq1_how_does_synthetic}

We first ask whether a CRS can be trained without any in-domain conversational corpus by converting non-conversational domain signals into synthetic conversational supervision.

We compare popularity ranking, zero-shot prompting, na\"ive GPT-generated training data, and our domain-grounded synthetic data across Qwen2.5-1.5B, Qwen3-4B, Llama3-3B, and NBCRS. For LLM backbones, we evaluate both LoRA and Full-SFT when applicable; for NBCRS, we compare raw seed data against the synthesized conversational data. \Cref{tab:rq1_combined_recall} reports Recall@$k$ on ReDial and INSPIRED.

\begin{table*}[htb]
\small
\setlength{\tabcolsep}{3pt}
\renewcommand{\arraystretch}{0.88}
\centering
\resizebox{\linewidth}{!}{%
\begin{tabular}{c|c|cccc|cccc}
\toprule
\multirow{2}{*}{Model} & \multirow{2}{*}{Setting}
  & \multicolumn{4}{c|}{ReDial} & \multicolumn{4}{c}{INSPIRED} \\
& & R@1 & R@5 & R@10 & R@20 & R@1 & R@5 & R@10 & R@20 \\
\midrule
\multirow{2}{*}{\textbf{Popularity}}
 & Seed Dataset
   & 0.09\se{0.04} & 0.24\se{0.06} & 0.33\se{0.07} & 0.34\se{0.07}
   & 0.03\se{0.03} & 0.03\se{0.03} & 0.55\se{0.13} & 0.70\se{0.15} \\
 & Synth Data
   & \textbf{0.56}\se{0.09} & \textbf{1.87}\se{0.16} & \textbf{2.46}\se{0.19} & \textbf{3.72}\se{0.23}
   & \textbf{0.30}\se{0.10} & \textbf{0.85}\se{0.16} & \textbf{1.77}\se{0.23} & \textbf{2.35}\se{0.26} \\
\midrule
\multirow{4}{*}{\textbf{Qwen2.5-1.5B}}
 & Zero-shot
   & 2.39\se{0.19} & 6.96\se{0.31} & 8.15\se{0.33} & 9.51\se{0.36}
   & 0.64\se{0.14} & 1.43\se{0.21} & 2.01\se{0.24} & 2.35\se{0.26} \\
 & GPT-Generated
   & 2.56\se{0.19} & 7.63\se{0.32} & 9.29\se{0.35} & 10.71\se{0.38}
   & 0.76\se{0.17} & 1.89\se{0.26} & 2.34\se{0.29} & 3.29\se{0.35} \\
 & \cellcolor{gray!15}Synth-LoRA
   & \cellcolor{gray!15}\textbf{3.39}\se{0.22} & \cellcolor{gray!15}9.27\se{0.35} & \cellcolor{gray!15}11.13\se{0.38} & \cellcolor{gray!15}12.83\se{0.41}
   & \cellcolor{gray!15}1.17\se{0.21} & \cellcolor{gray!15}3.21\se{0.34} & \cellcolor{gray!15}4.43\se{0.40} & \cellcolor{gray!15}5.82\se{0.46} \\
 & \cellcolor{gray!25}Synth-SFT
   & \cellcolor{gray!25}3.36\se{0.22} & \cellcolor{gray!25}\textbf{9.38}\se{0.36} & \cellcolor{gray!25}\textbf{11.40}\se{0.39} & \cellcolor{gray!25}\textbf{13.29}\se{0.42}
   & \cellcolor{gray!25}\textbf{1.97}\se{0.27} & \cellcolor{gray!25}\textbf{4.58}\se{0.41} & \cellcolor{gray!25}\textbf{6.09}\se{0.47} & \cellcolor{gray!25}\textbf{7.90}\se{0.52} \\
\midrule
\multirow{4}{*}{\textbf{Qwen3-4B}}
 & Zero-shot
   & 7.09\se{0.31} & 19.58\se{0.48} & 22.70\se{0.51} & 26.87\se{0.54}
   & 5.42\se{0.40} & 10.20\se{0.53} & 12.03\se{0.57} & 13.80\se{0.60} \\
 & GPT-Generated
   & 7.45\se{0.32} & 22.19\se{0.51} & 25.62\se{0.53} & 28.74\se{0.55}
   & 5.94\se{0.46} & 11.12\se{0.61} & 12.86\se{0.65} & 14.03\se{0.68} \\
 & \cellcolor{gray!15}Synth-LoRA
   & \cellcolor{gray!15}\textbf{8.39}\se{0.34} & \cellcolor{gray!15}\textbf{23.72}\se{0.52} & \cellcolor{gray!15}26.26\se{0.54} & \cellcolor{gray!15}28.79\se{0.55}
   & \cellcolor{gray!15}5.82\se{0.46} & \cellcolor{gray!15}10.48\se{0.60} & \cellcolor{gray!15}12.07\se{0.63} & \cellcolor{gray!15}12.86\se{0.65} \\
 & \cellcolor{gray!25}Synth-SFT
   & \cellcolor{gray!25}7.73\se{0.33} & \cellcolor{gray!25}22.81\se{0.51} & \cellcolor{gray!25}\textbf{26.32}\se{0.54} & \cellcolor{gray!25}\textbf{29.80}\se{0.56}
   & \cellcolor{gray!25}\textbf{6.01}\se{0.46} & \cellcolor{gray!25}\textbf{11.35}\se{0.62} & \cellcolor{gray!25}\textbf{13.12}\se{0.66} & \cellcolor{gray!25}\textbf{14.67}\se{0.69} \\
\midrule
\multirow{4}{*}{\textbf{Llama3-3B}}
 & Zero-shot
   & 1.57\se{0.15} & 5.56\se{0.28} & 9.07\se{0.35} & 13.26\se{0.41}
   & 2.04\se{0.25} & 5.00\se{0.38} & 6.94\se{0.44} & 8.89\se{0.50} \\
 & GPT-Generated
   & 1.54\se{0.15} & 5.86\se{0.29} & 8.96\se{0.35} & 13.47\se{0.42}
   & 1.97\se{0.27} & 5.03\se{0.43} & 7.07\se{0.50} & 9.34\se{0.57} \\
 & \cellcolor{gray!15}Synth-LoRA
   & \cellcolor{gray!15}1.53\se{0.15} & \cellcolor{gray!15}5.71\se{0.28} & \cellcolor{gray!15}8.79\se{0.35} & \cellcolor{gray!15}13.35\se{0.42}
   & \cellcolor{gray!15}1.97\se{0.27} & \cellcolor{gray!15}5.07\se{0.43} & \cellcolor{gray!15}7.26\se{0.50} & \cellcolor{gray!15}9.42\se{0.57} \\
 & \cellcolor{gray!25}Synth-SFT
   & \cellcolor{gray!25}\textbf{1.91}\se{0.17} & \cellcolor{gray!25}\textbf{6.54}\se{0.30} & \cellcolor{gray!25}\textbf{9.72}\se{0.36} & \cellcolor{gray!25}\textbf{14.42}\se{0.43}
   & \cellcolor{gray!25}\textbf{2.61}\se{0.31} & \cellcolor{gray!25}\textbf{5.75}\se{0.45} & \cellcolor{gray!25}\textbf{7.83}\se{0.52} & \cellcolor{gray!25}\textbf{9.98}\se{0.58} \\
\midrule
\multirow{2}{*}{\textbf{NBCRS}}
 & Seed Dataset
   & 0.00\se{0.00} & 0.39\se{0.28} & 0.78\se{0.39} & 1.17\se{0.48}
   & 0.00\se{0.00} & 0.00\se{0.00} & 0.00\se{0.00} & 0.00\se{0.00} \\
 & Synth Data
   & \textbf{2.54}\se{0.70} & \textbf{11.52}\se{1.41} & \textbf{19.53}\se{1.75} & \textbf{27.34}\se{1.97}
   & \textbf{4.55}\se{2.23} & \textbf{9.09}\se{3.08} & \textbf{13.64}\se{3.68} & \textbf{23.86}\se{4.57} \\
\bottomrule
\end{tabular}
}
\caption{RQ1 Recall (\%) results on ReDial and INSPIRED. Small gray values denote standard errors. Shaded entries denote LLM synthetic fine-tuning variants; bold marks the best Recall in each model block and dataset.}
\label{tab:rq1_combined_recall}
\end{table*}

\paragraph{Finding 1: Domain-grounded synthetic data consistently outperforms zero-shot and na\"ive generation.}
Models trained on domain-grounded synthetic data outperform both zero-shot baselines and models trained on GPT-generated data (produced without domain-specific seed signals) across backbone architectures and benchmarks.
On INSPIRED with Qwen2.5-1.5B, SFT-tuned models achieve +207.8\% Recall@1 improvement over zero-shot, compared to only +18.8\% for GPT-generated data.
This indicates that incorporating domain-specific signals during synthetic data generation significantly improves recommendation accuracy beyond what generic LLM generation provides.

\paragraph{Finding 2: Smaller models benefit disproportionately from synthetic fine-tuning.}
The relative gains from domain-grounded synthetic data are most pronounced for smaller models. Qwen2.5-1.5B sees Recall@1 improvements of +41.8\% (LoRA) and +40.6\% (SFT) on ReDial, while the larger Qwen3-4B achieves +18.3\% and +9.0\% respectively, suggesting that smaller models benefit more from domain-specific fine-tuning data.

\paragraph{Finding 3: Full-SFT generally outperforms LoRA, particularly for smaller models.}
Across most configurations, Full-SFT yields stronger results than LoRA-based fine-tuning, with the gap being especially pronounced for smaller models (e.g., Qwen2.5-1.5B on INSPIRED: +207.8\% vs.\ +82.8\% for Recall@1). For Llama3-3B, LoRA fine-tuning sometimes underperforms zero-shot, while Full-SFT consistently improves. This suggests that when training data is synthetically generated, full parameter updates are often important for adapting the model sufficiently.

\paragraph{Finding 4: The bootstrapping pipeline generalizes beyond LLM-based CRS.}
NBCRS~\citep{xie2024neighborhood}, a traditional neighborhood-based CRS, achieves substantially higher performance when trained on synthesized conversational data compared to raw seed data (e.g., Recall@5 improves from 0.39 to 11.52 on ReDial). This confirms that the data transformation---from non-conversational signals to conversational format---provides value independent of the downstream model architecture.

\paragraph{Synthetic-data quality.} We separately assess the generated corpus itself: human ratings of sampled query-recommendation pairs average close to 4 on a 1--5 scale across naturalness, coherence, relevance, and diversity, and both corpora follow the requested output format. Notably, the na\"ive corpus matches the item catalog \emph{more} often than the grounded corpus while performing worse downstream, indicating that catalog overlap does not track the usefulness of the supervision. Full human-evaluation and title-validity results are reported in \Cref{app:data_quality}.

\paragraph{Takeaway 1.}
The decisive factor is not simply using an LLM to generate more text, but grounding generation in domain-specific seed signals. This supports zero-data bootstrapping as a practical alternative when conversational logs are unavailable.

\subsection{RQ2: Active Selection Improves Data Efficiency}
\label{sec:rq2_how_does_active}

Synthetic data generation incurs teacher-LLM cost, so the next question is whether active selection can identify more useful seed items under fixed generation budgets.

We compare Jensen-Shannon (JS) divergence, Fisher information, random sampling, and a popularity heuristic that selects items with the highest review counts. We fine-tune Llama3-3B and Qwen2.5-1.5B on data generated from each selected subset and evaluate Recall@$k$ ($k\in\{1,5\}$) on ReDial and INSPIRED.

\begin{figure}[htb]
    \centering
    \includegraphics[width=\columnwidth]{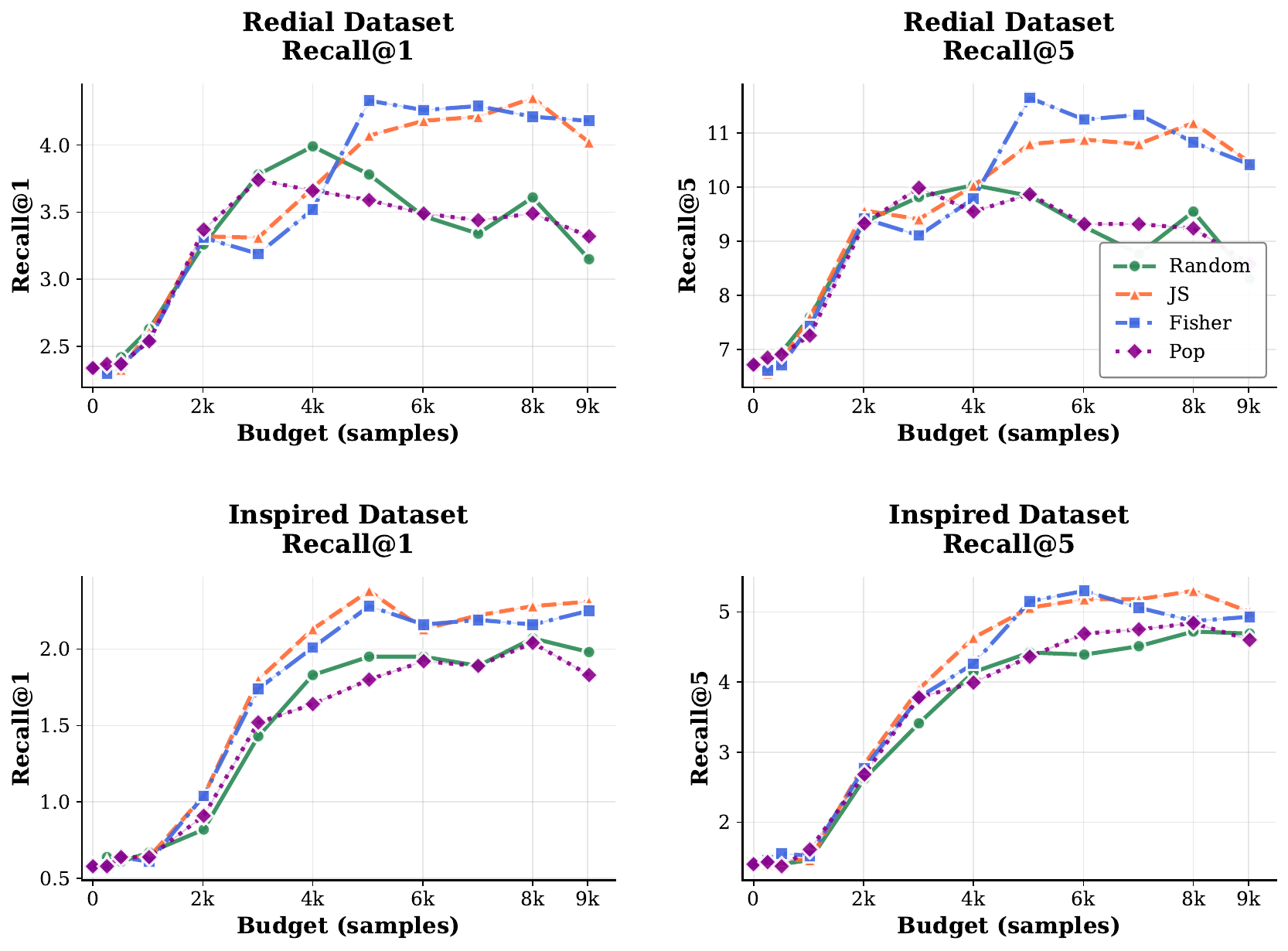}
    \caption{RQ2: Qwen2.5-1.5B performance across budgets on ReDial and INSPIRED. Active selection strategies (JS, Fisher) consistently outperform random sampling and popularity-based selection.}
    \label{fig:qwen2.5_metrics}
\end{figure}

\Cref{fig:qwen2.5_metrics} and \Cref{fig:llama3_metrics} compare performance across budgets.

\paragraph{Finding 5: Both active strategies outperform random sampling and popularity-based selection, but the margin depends on budget and signal.}
JS and Fisher consistently identify more beneficial seed items than random or popularity-based selection, reducing teacher-LLM calls while improving performance at similar budgets. Popularity, which prioritizes high-review-count items, often underperforms even random sampling, consistent with prior observations that random sampling can be a strong baseline in text-generation tasks~\citep{perlitz2023active}.

\paragraph{Finding 6: JS and Fisher exhibit complementary strengths.}
JS prioritizes distributional coverage, while Fisher targets parameter-level information gain aligned with the fine-tuning objective~\citep{kirsch2022unifyingapproachesactivelearning}. Their relative advantage varies by dataset and budget, confirming that they capture different facets of sample informativeness and that domain popularity alone is not a reliable proxy.
The same ordering holds under NDCG at an equal teacher-call budget (\Cref{app:ndcg}), so active selection improves the rank position of correct items and not only their presence in the list. We further analyze the popularity profile and recommendation coverage of active selection in \Cref{app:supplementary_analysis}.

\paragraph{Takeaway 2.}
For zero-data CRS, seed selection should be treated as part of the data-generation algorithm rather than as a preprocessing detail. Information-theoretic criteria make teacher queries more efficient than simple item-frequency heuristics.

\subsection{RQ3: Domain Signals Improve Active Selection}
\label{sec:rq3_domain_signals}

Non-conversational domains provide multiple signal types. We ask whether metadata and collaborative filtering signals improve selection beyond review text alone. We isolate domain-signal effects by comparing semantic-only strategies (\textit{JS}, \textit{Fisher}), metadata-aware strategies (\textit{JS+Meta}, \textit{Fisher+Meta}), collaborative-filtering-aware strategies (\textit{JS+CF}, \textit{Fisher+CF}), and \textit{Random}. Unless otherwise noted, this analysis uses Llama3 across budget levels on ReDial and INSPIRED.

\paragraph{Metadata Provides Structural Coverage.}

\begin{figure}[htb]
    \centering
    \includegraphics[width=\columnwidth]{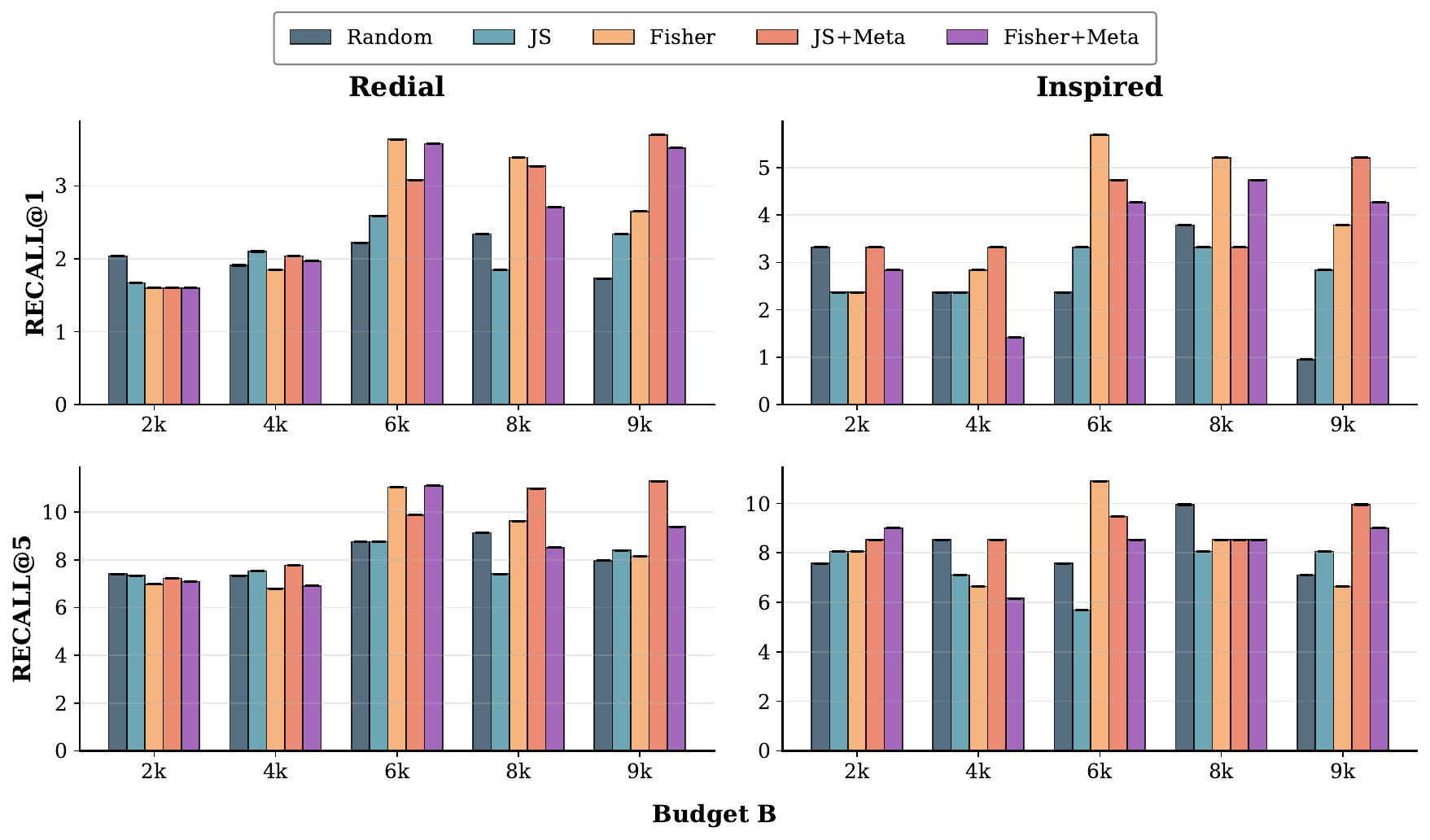}
    \caption{RQ3: Metadata-aware selection improves Recall@1 and Recall@5 over semantic-only and random selection, especially at larger budgets.}
    \label{fig:rq3}
\end{figure}

\Cref{fig:rq3} shows Recall@$1$ and Recall@$5$ across budget levels for both datasets using Llama3.

\paragraph{Finding 7: Metadata-aware selection consistently improves performance, with gains most pronounced at larger budgets.}
Methods incorporating item metadata outperform semantic-only and random baselines across both datasets. The cumulative advantage grows with budget, suggesting that metadata enables active selection to sustain informative sample identification over longer selection horizons. This confirms that structural domain signals (descriptions, genres, categories) provide context beyond what review text alone captures.

\paragraph{Collaborative Signals Add Preference Structure.}

\begin{figure}[htb]
    \centering
    \includegraphics[width=\columnwidth]{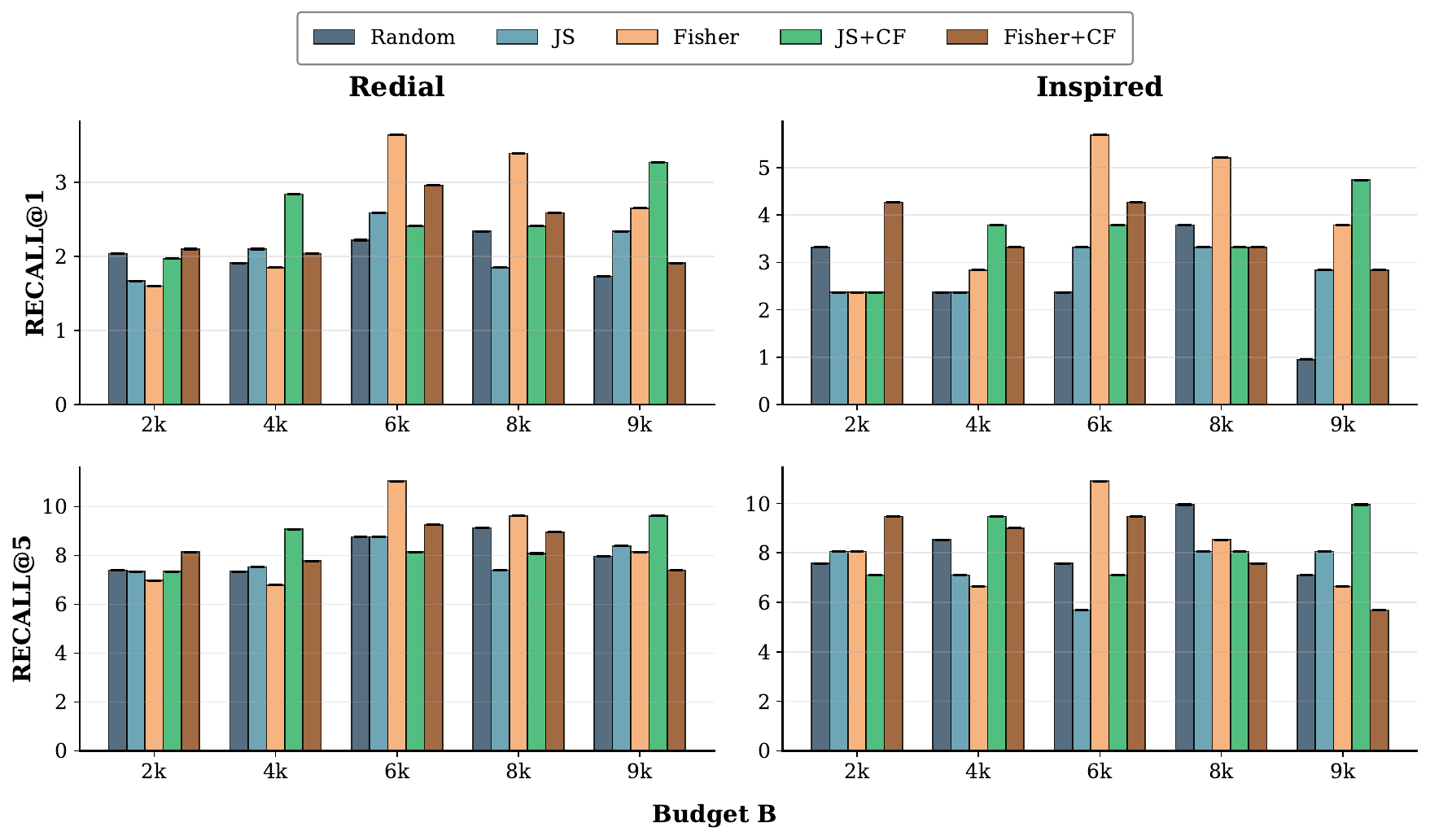}
    \caption{RQ3: Collaborative filtering signals provide complementary gains to semantic and metadata-based selection.}
    \label{fig:rq3_cf}
\end{figure}

\Cref{fig:rq3_cf} presents results across budget levels for both datasets.

\paragraph{Finding 8: Collaborative filtering signals provide additional gains orthogonal to semantic features.}
Collaborative-aware methods consistently demonstrate improved recommendation accuracy over baselines, with noticeable gains at larger budgets. This suggests that relational signals, which encode user preference patterns from interaction histories, enrich the selection process by emphasizing realistic, user-centric interactions that complement the content-based information in reviews and metadata.

\paragraph{Takeaway 3.}
Domain signals should be used not only as prompt context for generation, but also as selection signals. Metadata offers an easily available structural signal, while collaborative histories add preference information when such logs exist.

\subsection{RQ4: Synthetic Data Complements Scarce Real Data}
\label{sec:rq4_how_does_training}

\begin{figure}[htb]
    \centering
    \includegraphics[width=\columnwidth]{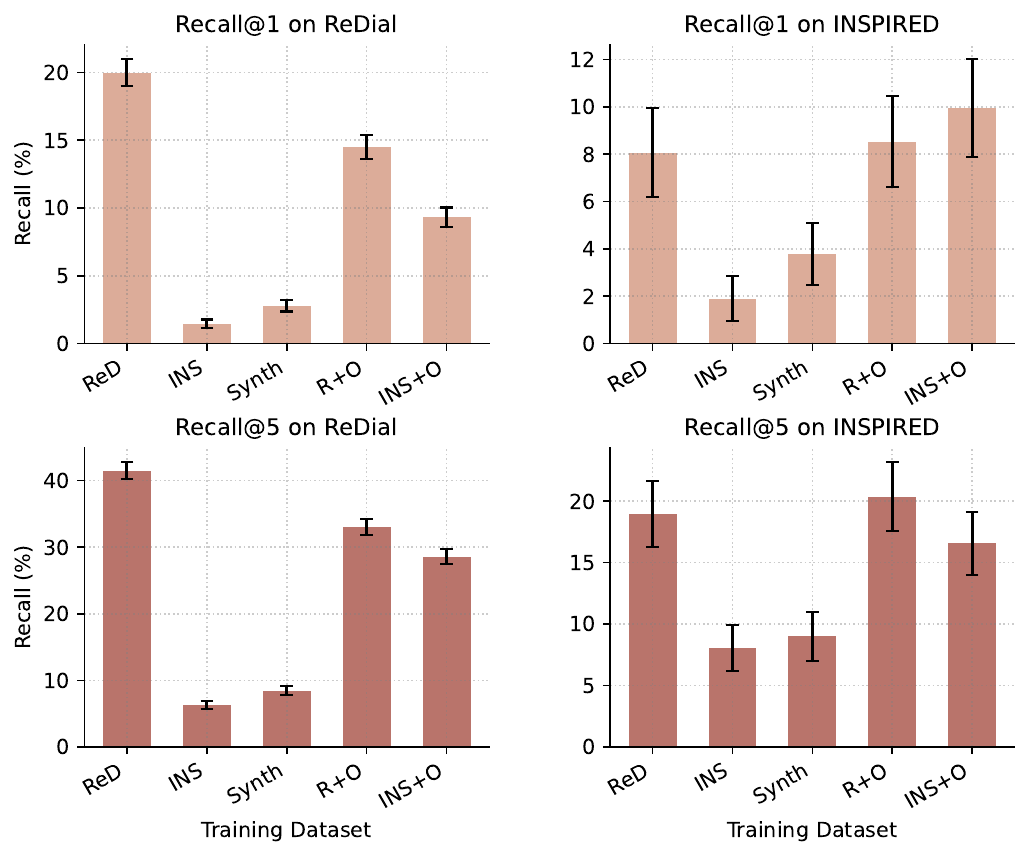}
    \caption{RQ4: Synthetic data is most beneficial when real conversational data is scarce and becomes less helpful when in-domain dialogue coverage is already large.}
    \label{fig:rq4_training}
\end{figure}

Finally, we ask whether synthetic data substitutes for or complements scarce real conversational data.
We compare models trained on \textbf{ReD} (ReDial only), \textbf{INS} (INSPIRED only), \textbf{Synth} (synthetic data only), \textbf{R+O} (ReDial + our synthetic data), and \textbf{INS+O} (INSPIRED + our synthetic data). \Cref{fig:rq4_training} compares these training mixtures.

\paragraph{Finding 9: In low-resource settings, synthetic data alone outperforms real in-domain data.}
On INSPIRED (${\sim}$1k dialogues), the model trained solely on synthetic data (\textbf{Synth}) outperforms the model trained on the original INSPIRED training set (\textbf{INS}), showing that domain-grounded synthetic data generated from non-conversational signals can surpass actual in-domain conversational data when the latter is scarce.

\paragraph{Finding 10: Combining synthetic and real data yields further gains, confirming complementarity.}
On INSPIRED, training with synthetic data alongside scarce or related real dialogues (\textbf{INS+O} and \textbf{R+O}) substantially improves over \textbf{INS} alone. On ReDial, \textbf{ReD} remains competitive and \textbf{R+O} slightly reduces performance, suggesting that synthetic augmentation is most valuable when real conversational coverage is limited.

\paragraph{Takeaway 4.}
Synthetic bootstrapping is most useful precisely in the low-resource setting that motivates zero-data CRS. When real conversational coverage is already broad, synthetic data should be used more selectively or filtered more aggressively.

\section{Related Work}
\paragraph{Conversational Recommender Systems.}
CRS supports interactive recommendation, often with mixed initiative between user and system~\citep{christakopoulou2016towards, li2018towards, chen-etal-2019-towards, lei2020estimation,he2022bundle, zhang2022multiple, xia2023user}.
Recent work improves CRS by integrating recommendation and conversation modules, incorporating language models, or directly prompting and fine-tuning LLMs for recommendation~\citep{wang22unicrs, xie2024neighborhood,friedman2023leveraging, yang2024unleashing,he2023large, feng2023large,xi2024memocrs}.
Our study targets a stricter setting: fine-tuning LLM-based CRS when \emph{no} in-domain conversational data is available, using only non-conversational domain signals.

\paragraph{Data Augmentation for CRS.}
Data scarcity is a central challenge for CRS, where cold-start issues are common and large-scale conversational interactions are difficult to obtain~\citep{chae2020ar,mysore2023large,leszczynski2023talk,wang2023improving}.
Recent augmentation methods use LLMs to simulate conversations involving long-tail items~\citep{zhao2023alleviating}, generate synthetic narrative queries from user-item histories~\citep{mysore2023large}, synthesize music-domain conversations from item collections~\citep{leszczynski2023talk}, or enrich existing dialogues with retrieved items, hard negatives, and neighboring conversations~\citep{jandaghi2024faithful,xu-etal-2025-beyond,10.1145/3712588,wang2023improving}.
These methods still operate \emph{on top of} existing in-domain conversations. LLM-REDIAL~\citep{liang2024llmredial} instead verbalizes user histories into a large multi-domain synthetic corpus, complementary to our focus on which non-conversational signals and selection strategies yield the most useful supervision under a fixed teacher budget. Our zero-data setting bootstraps conversational training data entirely from reviews, metadata, and collaborative interactions, without any pre-existing conversational corpus.

\paragraph{Active Learning.}
Active learning has been used in recommender systems and NLP to maximize knowledge transfer from LLMs while minimizing costly queries~\citep{du2024active,zhang2024elad, liu2024evolving,di2024performance,li2024enhancing,xia2025selection}.
Prior work selects informative user sessions for teacher annotation~\citep{du2024active}, generates targeted examples from student errors~\citep{liu2024evolving}, or guides query selection with model explanations and teacher corrections~\citep{zhang2024elad}.
We apply established information-theoretic criteria (Jensen-Shannon divergence and Fisher information) to a different problem: selecting non-conversational seed instances for synthetic CRS data generation. Our contribution is not a new active learning criterion, but a systematic empirical study of how active selection interacts with CRS-specific domain signals in a zero-data pipeline.

\section{Conclusion}

We presented a systematic empirical study of zero-data CRS bootstrapping: generating conversational supervision from non-conversational domain signals via LLMs. Across models, benchmarks, and fine-tuning paradigms, domain-grounded synthetic data outperforms zero-shot and na\"ive synthetic baselines; active selection improves data efficiency; metadata and collaborative signals strengthen selection; and synthetic data is especially effective when real dialogues are scarce. These findings show that non-conversational signals, when systematically selected and transformed, offer a practical path to building CRS without any conversational training data.

\paragraph{LLM Usage:}
We used LLMs solely for grammar refinement and wording edits in drafting parts of this paper.

\section*{Limitations}

We evaluate exclusively on the movie domain. The selection and bootstrapping mechanics are domain-general, but transfer to domains such as music or e-commerce would require domain-specific seed corpora, generation templates, and evaluation benchmarks; establishing cross-domain generalization is therefore left to future work. The bootstrapping pipeline relies on LLM-generated synthetic data, which may contain hallucinations or stylistic artifacts that propagate to downstream models. Finally, the quality of generated dialogues depends on prompt design. While recent work on prompt optimization could potentially improve generation quality, we do not explore this direction.

\appendix
\section{Selection Criteria Details}
\label{app:selection_criteria}

This appendix provides the construction details for the two active selection criteria defined in \Cref{sec:selection_strategies}.

\paragraph{JS Diversity.}
We partition the embedding space using $K$-means clustering to obtain cluster centers $C = \{\mathbf{c}_1, \ldots, \mathbf{c}_K\} \subset \mathbb{R}^d$. For each embedding $\mathbf{x}_i$, we compute its Euclidean distance to each cluster center~\citep{walkowiak2019evaluation,ikotun2023k}, $d(\mathbf{x}_i, \mathbf{c}_j)=\|\mathbf{x}_i-\mathbf{c}_j\|$, and construct a soft cluster-membership distribution $p_i=\{p_{ij}\}_{j=1}^K$ via softmax:
\begin{equation}
\nonumber
    p_{ij} = \frac{\exp\bigl(-d(\mathbf{x}_i, \mathbf{c}_j)\bigr)}{\sum_{k=1}^{K} \exp\bigl(-d(\mathbf{x}_i, \mathbf{c}_k)\bigr)}, \quad j = 1, \ldots, K.
\end{equation}
High entropy $H(p_i)$ indicates that $\mathbf{x}_i$ lies close to multiple centroids, while low entropy indicates strong association with a single cluster. We initialize $\mathcal{R}_0=\{i^{(1)}\}$ with the maximum-entropy sample~\citep{shannon1948mathematical}:
\begin{equation}
\nonumber
i^{(1)}=\arg\max_{i\in\mathcal{D}} H(p_i),
\quad
H(p_i)=-\sum_{j=1}^{K} p_{ij}\log p_{ij}.
\end{equation}
Subsequent samples are selected by \Cref{eq:js_selection}. The entropy term favors boundary samples, while the JS term~\citep{lin2002divergence} promotes novelty relative to the selected set. We update $\mathcal{R}_{t+1}=\mathcal{R}_t\cup\{i^{(t)}\}$ and repeat until $|\mathcal{R}|=B$.

\paragraph{Fisher Information.}
Under a last-layer linearization of the fine-tuning objective~\citep{mukherjee2024optimal,kveton2025active}, greedy log-determinant selection approximates expected information gain~\citep{kirsch2022unifyingapproachesactivelearning}. We initialize $\Sigma_0=I_d$ and $\Lambda_0=I_d$. Given selected set $\mathcal{R}_t$, the regularized information matrix is
$
    \Sigma_t = I_d + \sum_{j \in \mathcal{R}_t} \mathbf{x}_j \mathbf{x}_j^\top,
    \Lambda_t = \Sigma_t^{-1}.
$
The next sample is selected by \Cref{eq:fisher_selection}, corresponding to the largest incremental information gain~\citep{doi:10.1137/1.9780898719109,ash2021gone,IMM2012-03274}. After selecting $i^{(t)}$, we update $\mathcal{R}_{t+1}=\mathcal{R}_t\cup\{i^{(t)}\}$ and update $\Lambda_t$ using the rank-$1$ Sherman--Morrison formula~\citep{10.1214/aoms/1177729893}:
\begin{equation}
\nonumber
    \Lambda_{t+1} \leftarrow \Lambda_t -
    \frac{(\Lambda_t \mathbf{x}_{i^{(t)}}) (\Lambda_t \mathbf{x}_{i^{(t)}})^\top}
    {1 + \mathbf{x}_{i^{(t)}}^\top \Lambda_t \mathbf{x}_{i^{(t)}}}.
\end{equation}
The Fisher criterion is equivalent to greedy D-optimal design~\citep{doi:10.1137/1.9780898719109}: each step maximizes $\log\det(\Sigma_{t+1})-\log\det(\Sigma_t)$, reducing the volume of the confidence ellipsoid for model parameters under the last-layer approximation. The JS criterion is motivated by submodular coverage objectives~\citep{10.1007/BF01588971}. Together, JS and Fisher represent complementary notions of sample utility: distributional coverage and parameter-aligned informativeness.

\section{Popularity Bias and Coverage Analysis}
\label{app:supplementary_analysis}

We further characterize the active-selection results in RQ2 by analyzing the popularity profile of selected seed items and the coverage of the resulting recommendations. These diagnostics show that information-theoretic selection identifies useful seeds across the item distribution and produces broader recommendation coverage.

\subsection{Selected-Item Popularity}
\label{app:popularity_analysis}

We partition seed items by Amazon review count into \emph{head} (top 20\%), \emph{mid} (next 30\%), and \emph{tail} (bottom 50\%) buckets. For budgets $B \in \{1\text{k}, 2\text{k}, 3\text{k}\}$, we report selected-bucket proportions, the Gini coefficient of the selected-item distribution, and the Spearman correlation $\rho$ between item popularity and selection probability.

\begin{table}[htb]
\small
\centering
\begin{tabular}{clccccc}
\toprule
$B$ & Method & Head & Mid & Tail & Gini & $\rho$ \\
\midrule
\multirow{3}{*}{1k}
 & Random & 20.6 & 32.5 & 46.9 & .706 & $+$.002 \\
 & JS     & 22.7 & 32.6 & 44.7 & .677 & $+$.042 \\
 & Fisher & 14.8 & 23.3 & 61.9 & .787 & $-$.172 \\
\midrule
\multirow{3}{*}{2k}
 & Random & 20.7 & 31.5 & 47.9 & .702 & $+$.001 \\
 & JS     & 22.7 & 32.8 & 44.5 & .684 & $+$.065 \\
 & Fisher & 18.2 & 26.1 & 55.8 & .763 & $-$.154 \\
\midrule
\multirow{3}{*}{3k}
 & Random & 20.4 & 32.0 & 47.6 & .717 & $+$.006 \\
 & JS     & 22.5 & 32.2 & 45.3 & .700 & $+$.076 \\
 & Fisher & 18.4 & 27.8 & 53.8 & .763 & $-$.145 \\
\bottomrule
\end{tabular}
\caption{Popularity composition of selected samples at budgets $B \in \{1\text{k}, 2\text{k}, 3\text{k}\}$ (Qwen2.5-1.5B). Head/Mid/Tail denote selected-item percentages by Amazon review-count bucket. Lower Gini indicates more equitable coverage; $\rho$ is the Spearman correlation between item popularity and selection probability.}
\label{tab:selection_bias}
\end{table}

\Cref{tab:selection_bias} shows a clear distinction between popularity-based selection and the active criteria. Fisher has a strong long-tail orientation: tail items account for 53.8--61.9\% of selected samples, and selection probability is negatively correlated with review-count popularity across all budgets ($\rho=-.172$ to $-.145$). JS remains closer to the seed-pool distribution while maintaining only a mild positive popularity correlation ($\rho=+.042$ to $+.076$). Thus, the active strategies select useful seeds from broader regions of the item space, with Fisher especially highlighting informative long-tail items. Because these statistics are computed from the selected set alone, before any teacher calls, they offer a low-cost check on selection quality: a concentrated or low-coverage selection can be corrected by falling back to uniform-random sampling, which RQ2 shows remains a usable baseline at every budget.

\subsection{Recommendation Coverage}
\label{app:stratified_recall}

We next evaluate how these selection patterns translate into model outputs. We bucket test-set ground-truth items by evaluation frequency and report Recall@5 for each bucket using Qwen2.5-1.5B. We also report Coverage@$K$, the number of unique items appearing in top-$K$ recommendations across all test conversations.

\begin{table}[htb]
\small
\centering
\begin{tabular}{lcccccc}
\toprule
& \multicolumn{3}{c}{ReDial R@5} & \multicolumn{3}{c}{INSPIRED R@5} \\
\cmidrule(lr){2-4} \cmidrule(lr){5-7}
Method & Head & Mid & Tail & Head & Mid & Tail \\
\midrule
Random & 9.44 & 5.58 & 3.70 & 5.47 & 3.83 & 3.06 \\
JS     & 11.37 & 7.44 & 3.89 & 5.89 & 4.94 & 3.06 \\
Fisher & 12.10 & 7.54 & 4.44 & 6.25 & 4.78 & 2.64 \\
\bottomrule
\end{tabular}
\caption{Stratified Recall@5 by item popularity bucket on ReDial and INSPIRED (Qwen2.5-1.5B). Items are partitioned into head, mid, and tail buckets by evaluation-set frequency.}
\label{tab:stratified_recall}
\end{table}

\begin{table}[htb]
\small
\centering
\begin{tabular}{lcccc}
\toprule
& \multicolumn{2}{c}{ReDial} & \multicolumn{2}{c}{INSPIRED} \\
\cmidrule(lr){2-3} \cmidrule(lr){4-5}
Method & Cov@5 & Cov@10 & Cov@5 & Cov@10 \\
\midrule
Random & 1030 & 1407 & 502 & 696 \\
JS     & 1164 & 1571 & 569 & 802 \\
Fisher & 1188 & 1570 & 547 & 777 \\
\bottomrule
\end{tabular}
\caption{Coverage@$K$: number of unique items in top-$K$ recommendations across all test conversations (Qwen2.5-1.5B). Higher values indicate greater recommendation diversity.}
\label{tab:coverage}
\end{table}

On ReDial, both active strategies improve Recall@5 over Random across head, mid, and tail buckets (\Cref{tab:stratified_recall}); Fisher raises tail Recall@5 from 3.70 to 4.44 while also improving head Recall@5 from 9.44 to 12.10. On INSPIRED, JS improves head and mid performance while maintaining tail Recall@5, and Fisher gives the strongest head performance among the three methods. Coverage shows the most consistent pattern: JS and Fisher recommend substantially more unique items than Random on both datasets (\Cref{tab:coverage}). Overall, active selection improves both recommendation accuracy and coverage, reinforcing the RQ2 conclusion that information-theoretic seed selection is more effective than random sampling.

\section{Ranking Quality (NDCG)}
\label{app:ndcg}

\Cref{tab:ndcg} reports NDCG for the RQ2 selection strategies at the full generation budget (Qwen2.5-1.5B). We omit NDCG@1, which is identical to Recall@1 when each evaluation instance has a single ground-truth item.

\begin{table}[htb]
\small
\centering
\begin{tabular}{lcccccc}
\toprule
& \multicolumn{3}{c}{ReDial} & \multicolumn{3}{c}{INSPIRED} \\
\cmidrule(lr){2-4} \cmidrule(lr){5-7}
Method & @5 & @10 & @20 & @5 & @10 & @20 \\
\midrule
Random     & 5.82 & 6.53 & 7.03 & 3.41 & 3.85 & 4.27 \\
Popularity & 6.05 & 6.75 & 7.27 & 3.28 & 3.73 & 4.10 \\
JS         & 7.40 & 8.13 & 8.62 & \textbf{3.76} & \textbf{4.30} & \textbf{4.71} \\
Fisher     & \textbf{7.47} & \textbf{8.19} & \textbf{8.70} & 3.68 & 4.25 & 4.59 \\
\bottomrule
\end{tabular}
\caption{NDCG@$k$ (\%) at the full generation budget (Qwen2.5-1.5B). Both active criteria outperform random and popularity-based selection at every cutoff on both benchmarks.}
\label{tab:ndcg}
\end{table}

NDCG tracks Recall closely: both active criteria improve over random and popularity selection at every cutoff on both benchmarks, and their advantage varies with budget in the same way as the Recall curves in \Cref{fig:qwen2.5_metrics}. Active selection therefore improves the rank position of correct items, not only their presence in the list.

\section{Synthetic Data Quality}
\label{app:data_quality}

Because the pipeline depends on LLM generation, we assess the generated corpus directly along two axes: human judgments of query-recommendation quality, and automatic validation of output format and item validity.

\subsection{Human Evaluation}
\label{app:human_eval}

Two graduate students independently rated 100 randomly sampled synthetic query-recommendation pairs on a 1-5 scale along four dimensions: naturalness (does the query read like a real user request), coherence (is the request internally consistent), relevance (do the recommendations match the request), and diversity (is the list varied rather than repetitive). Agreement within one point is 83-90\% across dimensions.

\begin{table}[htb]
\small
\centering
\begin{tabular}{lc}
\toprule
Dimension & Mean score \\
\midrule
Naturalness & 3.92 \\
Coherence   & 3.94 \\
Relevance   & 3.79 \\
Diversity   & 4.18 \\
\bottomrule
\end{tabular}
\caption{Human evaluation of 100 sampled synthetic query-recommendation pairs (1-5 scale, two annotators). Averaging the two annotators, 84\% of the 400 example-dimension scores are at least 3.5.}
\label{tab:human_eval}
\end{table}

All four dimensions average close to 4, and averaging the two annotators, 84\% of the 400 example-dimension scores are at least 3.5. Relevance is the lowest-scoring dimension, consistent with the teacher occasionally returning loosely related items in the 20-item list.

\subsection{Title and Format Validation}
\label{app:title_validity}

We normalize every generated title (lowercasing, removing release years and punctuation) and check it against a verified catalog of 9{,}944 titles compiled from ReDial, INSPIRED, and the seed data.

\begin{table}[htb]
\small
\centering
\begin{tabular}{lccc}
\toprule
Corpus & In-catalog & Exactly 20 & Dup.\ lists \\
\midrule
Domain-grounded & 76.5 & 97.3 & 2.6 \\
Na\"ive GPT     & 92.3 & 100.0 & 1.3 \\
\bottomrule
\end{tabular}
\caption{Format and item validity (\%) of the generated corpora. In-catalog: generated titles matching the verified catalog. Exactly 20: lists with the requested number of items. Dup.\ lists: lists containing a repeated title.}
\label{tab:title_validity}
\end{table}

Both corpora follow the requested format almost always. Exact matching is a lower bound on validity, since long-tail, international, and alternatively formatted titles are absent from this small catalog. Notably, the na\"ive corpus matches the catalog \emph{more} often yet gains far less downstream (\Cref{tab:rq1_combined_recall}): catalog overlap rewards popular titles, not useful ones.

\section{Prompts and Implementation Details}
\label{app:prompts_impl}

\subsection{Implementation Details} \label{impl_det}
All experiments are implemented using PyTorch and conducted on NVIDIA A6000 GPUs with 48GB memory. Fine-tuning is a one-time cost of under four hours per run on a single A6000; at inference the fine-tuned model serves requests without further teacher calls. For training stability, we use the Adam optimizer coupled with a warm-up learning rate scheduler. The batch size is set to 32, with gradient accumulation steps of 8.

Training configurations are tailored to each research question: for RQ1 (synthetic data comparison), we use a learning rate of \(5 \times 10^{-6}\) and train for 5 epochs with early stopping to ensure convergence; for RQ2 (active-selection budget curves), we use a learning rate of \(5 \times 10^{-6}\) and train for 1 epoch; for RQ3 (domain-signal comparisons), we use a learning rate of \(5 \times 10^{-6}\) and train for 1 epoch; for RQ4 (real/synthetic training mixtures), we use a learning rate of \(1 \times 10^{-5}\) and train for 3 epochs. We synthesize conversational datasets following the procedure outlined in~\Cref{sec:synthetic_data}. Synthetic data is generated using OpenAI's GPT-4o API~\citep{hurst2024gpt}, with a generation temperature of 0.8. We fix $\lambda=0.5$ (\Cref{eq:js_selection}) in all experiments.

\subsection{Experiments}

\begin{figure}[htb]
\centering
    \includegraphics[width=\columnwidth]{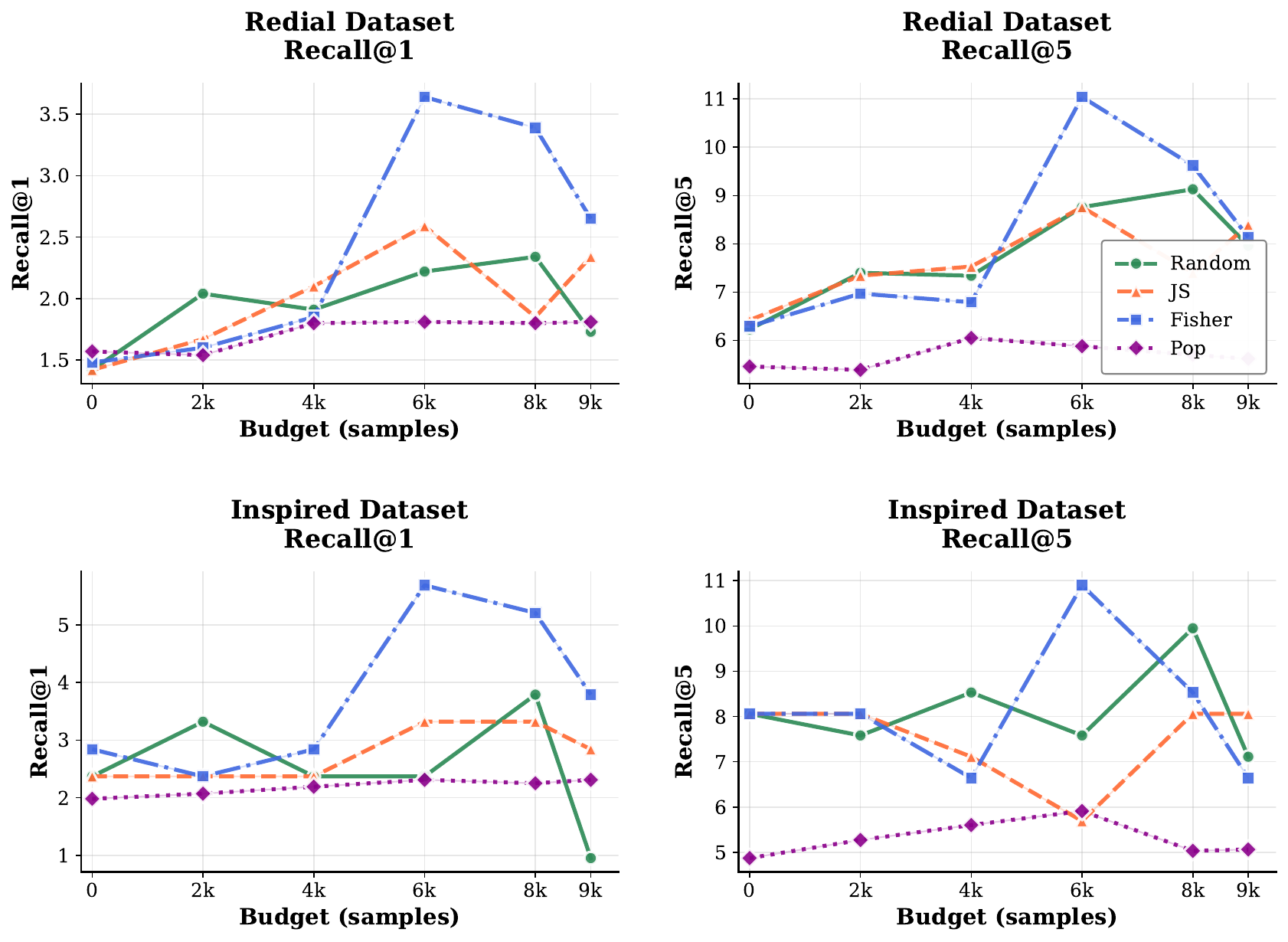}
    
    \caption{RQ2: Llama3-3B performance across budgets on ReDial and INSPIRED. Active selection strategies (JS, Fisher) consistently outperform random sampling and popularity-based selection.}
    \label{fig:llama3_metrics}
\end{figure}

\begin{queryboxenv}[title=Sample generated user-query and assistant response]\label{box:sample}

\textbf{User Query 1}\\
Movies with a unique twist on classic tales? I'm interested in films that take a well-known story and give it a fresh spin. I recently watched a Hercules movie that wasn't your typical mythological tale—it had humor and The Rock was pretty awesome, even if his acting chops are still a work in progress. Can you suggest other films like that? Maybe ones that add humor or a cool new perspective to a classic narrative?

\smallskip
\textbf{Assistant Response:} \textit{Shrek; Enchanted; Into the Woods; The Princess Bride; Stardust; 10 Things I Hate About You; Warm Bodies; Clueless; Ella Enchanted; Maleficent; The Lego Movie; Tangled; John Tucker Must Die; Easy A; The Many Adventures of Winnie the Pooh; Hook; Ever After: A Cinderella Story; The Nightmare Before Christmas; Pan; The Brothers Grimm.}

\medskip\hrule\medskip

\textbf{User Query 2}\\
Looking for a quirky, whimsical movie with a great ensemble cast and a bit of adventure. I've heard ``The Brothers Bloom'' might fit the bill with its delightful, offbeat charm and diverse settings. If you enjoy films like ``Dirty Rotten Scoundrels'' and appreciate a touch of the macabre mixed with a fun romp, this might be the one! Would love to know if there are other similar movies out there that capture this unique blend of humor and adventure. Any recommendations?

\smallskip
\textbf{Assistant Response:} \textit{The Grand Budapest Hotel; The Royal Tenenbaums; Moonrise Kingdom; Snatch; Amélie; Big Fish; Lock, Stock and Two Smoking Barrels; The Adventures of Tintin; In Bruges; The Life Aquatic with Steve Zissou; The Darjeeling Limited; Midnight in Paris; Scott Pilgrim vs.\ the World; The Men Who Stare at Goats; Kiss Kiss Bang Bang; The Fantastic Mr.\ Fox; Delicatessen; O Brother, Where Art Thou?; Catch Me If You Can; The Imaginarium of Doctor Parnassus.}

\medskip\hrule\medskip

\textbf{User Query 3}\\
My partner and I watched a film recently and thoroughly enjoyed it. However, I'm curious about movies where not much happens plot-wise, but dialogue and arguments drive the narrative. Any recommendations for films with this kind of storytelling approach? Also, any thoughts on movies that critique or play with horror fan stereotypes?

\smallskip
\textbf{Assistant Response:} \textit{My Dinner with Andre; Before Sunrise; The Breakfast Club; Clerks; Dazed and Confused; The Big Chill; Slacker, Richard Linklater's Waking Life; Coffee and Cigarettes; Frances Ha; Annie Hall; The Squid and the Whale; Glengarry Glen Ross; Tape; Carnage; What We Do in the Shadows; The Cabin in the Woods; Scream; Tucker and Dale vs. Evil; Behind the Mask: The Rise of Leslie Vernon.}

\end{queryboxenv}

\begin{queryboxenv}[title=Movie Recommendations Prompt]\label{box:resp}
{\ttfamily\small
Pretend you are a movie recommender system. I will give you a conversation 
between a user and you (a recommender system). Based on the conversation, 
you reply with 20 recommendations without extra sentences.\par\medskip
Here is the conversation: \{\textit{conv}\}
}
\end{queryboxenv}

\begin{queryboxenv}[title=Query Generation Prompt]\label{box:query}
{\ttfamily\small
You will be given multiple product reviews. Your task is to convert the
sentiment, issues, or features mentioned in these reviews into one
distinct query or question. I want you to style it similar to the following
queries.\par\medskip
\textbf{Sample Queries:} \{\textit{sample\_queries\_text}\}\par\medskip
Now, convert the following product reviews into one distinct query with a
similar style.\par\medskip
\textbf{Input Review:} \{\textit{product\_reviews\_text}\}
}
\end{queryboxenv}

\begin{queryboxenv}[title=Sample Query Templates ($\mathcal{Q}_{\text{tmp}}$)]\label{box:query_templates}
{\small
The following are example queries from $\mathcal{Q}_{\text{tmp}}$, sourced from Reddit movie recommendation discussions. These templates guide the style and tone of synthetic query generation.\par\medskip

\textit{``Sick with Covid, dealing with some serious cabin fever. Any good carefree films that might lighten the mood?''}\par\medskip

\textit{``Movies about the friends made along the way. I'm interested in movies where the main character continually meets and befriends new interesting characters. Maybe the main character is on a journey. Maybe the relationships with the new characters both begin and end during the story. Something like `Into the Wild' would be perfect.''}
}
\end{queryboxenv}

\end{document}